\documentclass[12pt]{article}
\usepackage{amsfonts}
\usepackage{amsmath}
\usepackage{amssymb}
\usepackage{graphicx}
\usepackage{chngpage}
\usepackage{multirow}
\usepackage{color}
\usepackage{harvard}
\usepackage{setspace}

\newcommand{\be}{\begin{eqnarray}}
\newcommand{\ee}{\end{eqnarray}}
\newcommand{\bi}{\begin{itemize}}
\newcommand{\ei}{\end{itemize}}

\begin{document}

\title{Econometric Models of Network Formation\thanks{I would like to thank Marcel Fafchamps, Seth Richards-Shubik, J\"{o}rg Stoye, Martin Weidner and a reviewer for comments on earlier drafts.  Financial support from the Economic and Social Research Council ESRC grant RES-589-28-0001 to the Centre for Microdata Methods and Practice and from ESRC Large Research Grant ES/P008909/1 is gratefully acknowledged.}
}
\author{\'Aureo de Paula\thanks{%
University College London, C\emph{e}MMAP and IFS, London, UK.
ORCID: 0000-0003-3611-3448. E-mail: \texttt{a.paula@ucl.ac.uk}} \\
University College London, CeMMAP and IFS}
\date{This Version: January, 2020}
\maketitle

\begin{abstract}
\noindent This article provides a selective review on the recent literature on econometric models of network formation.  The survey starts with a brief exposition on basic concepts and tools for the statistical description of networks.  I then offer a review of dyadic models, focussing on statistical models on pairs of nodes and describe several developments of interest to the econometrics literature.  The article also presents a discussion of non-dyadic models where link formation might be influenced by the presence or absence of additional links, which themselves are subject to similar influences.  This is related to the statistical literature on conditionally specified models and the econometrics of game theoretical models.   I close with a (non-exhaustive) discussion of potential areas for further development.
\newline
KEYWORDS: Network econometrics, dyadic models, strategic network formation models.
\end{abstract}

\newpage

\section{Introduction}

Tam\'{a}s Erd\'{e}lyi and John Cummings met in their first year of high school in Queens, New York City.  Sometime later, John met Doug Colvin, who had moved to the neighbourhood from Germany where his father was stationed with the US Army.  Together with a fourth neighbourhood friend, Jeffrey Hyman, they formed a band in 1974 and took the name of ``The Ramones.''  

Regardless of one's opinion on the music by Johnny (John), Joey (Jeffrey), Dee Dee (Doug) and Tommy (Tam\'{a}s) Ramone, their encounter was a matter of opportunity, talent and personal affinity.\footnote{The off-stage personal relationships were not harmonious for long.  Nevertheless, `` `I can see now how it was only natural that I would gravitate toward Tommy, Joey, and Johnny Ramone,' he wrote. `They were the obvious creeps of the neighbourhood \dots' '' quoting Dee Dee Ramone by ``The Curse of the Ramones'' (\emph{Rolling Stone}, May, 2016 retrieved from \texttt{https://www.rollingstone.com/culture/culture-news/the-curse-of-the-ramones-165741/}).}  Related factors inform the formation of networks in many other settings, from friendships in economic and/or social contexts, to the establishment of business relationships across firms and trading ties between countries.  These links influence and are facilitated by other outcomes that depend on how a particular group is connected.  For example, smoking behaviour among adolescents influences and is affected by whom they are related to.  Production decisions are similarly influenced by the set of clients and suppliers for a firm.  Especially with the advent and availability of data on networks from surveys and administrative sources, it is increasingly feasible and potentially relevant to provide a systematic assessment of drivers and correlates for the formation of such relationships.

In this article I review recent developments in the nascent literature on the econometrics of networks that provide a means to assess those drivers and correlates.  While an established literature exists in other fields, spanning social sciences, physics and statistics, I focus here on tools related to the field of economics.  Much of it builds and gets inspiration from previous developments in other fields, but some of it is quintessentially related to research in economics.  This is the case, for instance, of estimable models of strategic network formation.

I start with a brief exposition on basic concepts and tools for the (statistical) description of networks.  I then offer a review of dyadic models and describe several developments of interest to the econometrics literature and connections with related literatures (e.g., panel data).  This is followed by a discussion of non-dyadic models, where link formation might be influenced by the presence or absence of additional links, which themselves are subject to similar influences.  This is related to the statistical literature on conditionally specified models and the econometrics of game theoretical models.   As in those literatures, potential simultaneity issues are central to the identification and estimation of the quantities of interest and I explore some of those here as well. I close with a (non-exhaustive) discussion of potential areas for further development.

This review complements other available overviews within econometrics by \citeasnoun{Graham_AR15}, \citeasnoun{chandrasekhar2015}, \citeasnoun{DePaula2017} and \citeasnoun{graham2019}.  In particular, it follows closely (but expands upon) previous surveys in \citeasnoun{DePaula2017} and corresponding material in \citeasnoun{grahamdepaula2020}.

\section{Preliminary Foundations} \label{sec:prelim}

Networks are typically represented by graphs.  A graph $g$ is a pair of sets $(\mathcal{N}_g,\mathcal{E}_g)$ of \emph{nodes} (or \emph{vertices}) $\mathcal{N}_g$ and \emph{edges} (or \emph{links} or \emph{ties}) $\mathcal{E}_g$.  I will denote the number of elements in these sets by $|\mathcal{N}_g| \equiv N$ and $|\mathcal{E}_g|$, respectively.  Vertices here represent economic agents like individuals, households or firms.  An edge represents a link or connection between two nodes in $\mathcal{N}_g$.  A graph is undirected when $\mathcal{E}_g$ is the set of unordered pairs with elements in $\mathcal{N}_g$, say $\{i,j\}$ with $i,j \in \mathcal{N}_g$.  (I abstract away from self-links here.)  An example are (reciprocal) informal risk-sharing networks based on kinship or friendship (e.g., \citeasnoun{fafchampslund2003}).  To accomodate directional relationships, edges are best modeled as ordered pairs,  say $(i,j) \in \mathcal{N}_g \times \mathcal{N}_g$.  These graphs, known as directed graphs (or digraphs), are more adequate for handling relatioships that do not require reciprocity or for which direction carries a particular meaning, as in a supplier-client relationship in a production network (e.g., \citeasnoun{atalayetal2011}).  Generalizations allow for weighted links, perhaps representing distances between two individuals or the intensity of a particular relationship.   Such weights can be represented as a mapping from the space of pairs (unordered or ordered) into the real line.   

A common representation of a graph is through its $|\mathcal{N}_g| \times |\mathcal{N}_g|$ adjacency matrix $W$, where each line represents a different node.  The components of $W$ mark whether an edge between nodes $i$ and $j$ (or from $i$ to $j$ in a digraph) is present or not and possibly its weight (in weighted graphs).  The adjacency matrix allows one to translate combinatorial operations into linear algebraic ones and can be quite useful in several settings.  For an adjacency matrix $W$ to a simple unweighted graph (i.e., no self-links and at most one link between any pair of nodes), the $ij$ element of matrix $W^k, k \in \{ 1, \dots, N-1 \}$, for instance, produce the number paths of length $k$ between $i$ and $j$.  Two graphs are said to be isomorphic if their adjacency matrices can be obtained from each other, through multiplication by a permutation matrix, for example.  This corresponds to a relabeling of the vertices in the corresponding graphs.

\subsection{Vertex Features}

Various measures can then be defined to characterize a particular vertex in the graph, to relate two or more vertices on a graph, or to represent a global feature of the graph at hand.  In what follows I focus on simple, unweighted graphs for ease of exposition.  An important characteristic for a particular vertex $i$, for example, is the set of neighbours incident with that vertex in a graph $g$, denoted by $N_i(g)$.   The cardinality of this set is known as the ``degree'' of that node, and one can then talk about the relative frequency of degrees in a given graph as a whole.  (In directed graphs, one can further distinguish ``in-degrees'' and ``out-degrees'' relating to inward and outward edges from and to a given node.)  A ``dense'' graph, for instance, is one in which nodes display a lot of connections, and a common measure of density is the average degree divided by $|\mathcal{N}_g|-1$, which is the maximum number of possible links available to any given node.   It is common to define the (geodesic) distance between these two as the shortest path between those two nodes.  A graph is then said to be connected if the distance between any two vertices is finite (i.e., there is at least one path between those nodes).  

One can also define various measures to characterize the typical subnetwork structure around a given vertex.  For brevity, I only mention a basic taxonomy of such measures as specific definitions are available in most introductory texts on the subject (see, for example, the excellent overview in \citeasnoun{jackson2009}).  A network aspect of particular interest in social settings is the degree of ``clustering'' in the system, intuitively summarized by the propensity that two neighbours to a given node are also themselves directly linked.   Theoretically, for example, it may be easier for clustered individuals to coordinate on certain collective actions since clustering may facilitate common knowledge (\citeasnoun{chwe2000}) and different clustering metrics are available to quantify this feature in a network.  

Another feature of potential interest in economic and social networks is the degree of ``centrality'' of a given vertex, and various measures of centrality are also available.  Those aim at characterizing how important a given node is in comparison to the remaining nodes in $g$.  Aside from how connected a given vertex is (degree centrality) or how far on average a vertex is from any other vertex in the network (closeness centrality), one can also compute the betweenness centrality, illustrating how crucial a given node is in connecting individuals.  Another family of popular centrality measures summarises a node's centrality in reference to its neighbours centrality (more on this later).  The simplest of these measures is the eigenvector centrality (a.k.a. Gould's index of accessibility), corresponding to the dominant eigenvector of the adjacency matrix (\citeasnoun{gould1967}, \citeasnoun{bonacich1972}).  If an individual is central because she is connected with central individuals, this means that the (eigenvector) centrality vector $\mathbf{c}$ is such that $\mathbf{c} = W \mathbf{c}$.  It is thus the eigenvector related to a unit eigenvalue.  For a normalised adjacency matrix where all rows add up to one, this is guaranteed to exist and be unique if the network is (strongly) connected by the Perron-Frobenius Theorem.  Variations on this centrality measure include, for example, the Google's PageRank index (\citeasnoun{brinpage1998}).  Among the most popular metrics in this family were those proposed by \citeasnoun{katz1953}  and \citeasnoun{bonacich1987}.  The Katz centrality of a node $i$ can be motivated by ascribing a value of $\tilde \beta^k > 0$ to each connection reached by a walk of length $k$.  If one adds up the weights for each individual, one has a centrality measure for each individual given by the components of the vector $\tilde \beta W \mathbf{1} + \tilde \beta^2 W^2 \mathbf{1}  + \tilde \beta^3 W^3 \mathbf{1} + \dots$  If $\tilde \beta$ is below the reciprocal of $W$'s largest eigenvalue, we can write the above as $\tilde \beta (\mathbf{I} - \tilde \beta W)^{-1} W \mathbf{1}$, where $\tilde \beta$ is a small enough positive number.  The Bonacich centrality generalizes this formula to a two-parameter index defined by the vector $\alpha (\mathbf{I} - \tilde \beta W)^{-1} W \mathbf{1}$.  Such measures turn out to play an important role in the analysis of games and dissemination on networks (e.g., \citeasnoun{bcz2006} and the survey by \citeasnoun{zenou2015}).

\subsection{Random Graphs}

Letting $\mathcal{G}$ be a particular set of graphs, one can define a probability model on this sample space.  These models can and usually are indexed by features common to the graphs in $\mathcal{G}$, like the number of vertices and/or other motifs.  

Regardless of the generative model one has in mind, the initial examination of a network typically involves the characterisation of some of the features alluded to above, such as the links pattern among node pairs, triads, tetrads and $k$-tuples in general, usually referred to as a dyad or triad census in the first two cases.  Isomorphic graphs on the $k$-tuple of nodes are usually classified within the same equivalence class.   For example, in a dyad from a digraph, a pair with one link will be classified equivalently whether the link is from $i$ to $j$ or from $j$ to $i$.  Thus, if the network is directed, there are three patterns of subgraphs on pairs: no link, one unreciprocated link from one of the nodes to the other one or two reciprocated links.  In an undirected network two patterns exist for a pair of nodes: either those two vertices are connected or not.    In triads, an undirected network will feature four patterns: zero, one, two or three links.  A directed network will produce a richer set with 16 triad patterns.  The number of patterns grows rapidly with $k$ and, in a directed network, ``there are so many tetrad types (218) and pentad types (9608) that a $k$-subgraph census for $k \ge 4$ is often more cumbersome than the original sociomatrix.'' (\citeasnoun{Holland_Leinhardt_SM76}).\footnote{Expressions for the first two moments of such $k$-subgraph censuses appear in \citeasnoun{Holland_Leinhardt_SM76} and further asymptotic results under additional assumptions on the link formation probabilities feature in \citeasnoun{Bickel_et_al_AS11} and \citeasnoun{Bhattacharya_Bickel_AS15}.}

One of the early network models imposes a uniform probability on the class of graphs with a given number of nodes, $|\mathcal{N}_g|$, and a particular number of edges, $|\mathcal{E}_g|$, for $g \in \mathcal{G}$ (see \citeasnoun{erdosrenyi1959} and \citeasnoun{erdosrenyi1960}).  Another basic, canonical random graph model is one in which the edges between any two nodes follow an independent Bernoulli distribution with equal probability, say $p$.  For a large enough number of nodes and sufficiently small probability of link formation $p$, the degree distribution approaches a Poisson distribution, and the model is consequently known as the Poisson random-graph model.  This class of models appears in \citeasnoun{gilbert1959} and \citeasnoun{erdosrenyi1960} and has since been studied extensively.  While simple to characterise, it fails to reproduce important dependencies observed in social and economic networks.  One category of models that aims at a better representation of the regularities usually encountered in social systems involves models where nodes are incorporated into the graph sequentially and form ties more or less randomly.  One such model is the ``preferential attachment'' model (\citeasnoun{barabasialbert1999}), whereby the establishment of new links is more likely for higher-degree existing nodes, producing degree distributions with Pareto tails (as well as other regularities usually observed; see the presentation in \citeasnoun{jackson2009} or \citeasnoun{kolaczyk2009} for a more thorough exposition).

Another alternative is to rely on more general (static) random graph models that explicitly acknowledge the probabilistic dependencies in link formation.  \citeasnoun{frankstrauss1986}, for example, focus on random graphs where two (random) edges that do not share a vertex are conditionally independent given the other remaining (random) edges.  It reflects the intuition that ties are not independent of each other, but their dependency arises only through those who are directly involved in the connections in question.  This, and a homogeneity assumption (i.e., that all graphs that are the same up to a permutation of vertices have the same probability), delivers that
$$\mathbb{P}(G=g) \propto \exp\left(\alpha_0 t + \sum_{k=1} \alpha_k s_k \right),$$
where $t$ is the number of triangles (completely connected triples of vertices) and $s_k$ is the number of $k$-stars (tuples of $k+1$ vertices where one of the vertices has degree $k$ and the remaining ones have degree one).  (Notice that the Erd\"{o}s-R\'{e}nyi model is a specific case of the above model, where $\alpha_0=\alpha_2=\dots=\alpha_k=0$.)  This structure suggests a class of probabilistic models that reproduce the exponential functional form above even in cases where the exact properties used by Frank and Strauss do not hold.  Those models are such that
$\mathbb{P}(G=g) \propto \exp\left(\sum_{k=1}^p \alpha_k S_k(g) \right),$
where $S_k(g), k=1,\dots,p$ enumerate features of the graph $g$.  These would be characteristics like the number of edges, the number of triangles and possibly many others.  These models are known as exponential random graph models (ERGMs) (or $p^*$ models in the social sciences literature, see \citeasnoun{rpkl2007}) and can be extended beyond undirected random graphs.  The models above form an exponential family of distribution over (random) graphs and exponential distributions (e.g., Bernoulli, Poisson) have well-known probabilistic and statistical properties.  For example, the vector $(S_1(g),\dots,S_p(g))$ constitute a $p$-dimensional sufficient statistic for the parameters $(\alpha_1,\dots,\alpha_p)$.  

Some recent articles in econometrics are closely related to the exponential random graph model above.  More generally, all the models above (and many others) are presented in detail elsewhere (e.g., \citeasnoun{bollobas2001}, \citeasnoun{jackson2009}, \citeasnoun{kolaczyk2009}) and I will selectively discuss features and difficulties as they arise in the literature reviewed here.  

\section{Dyadic Models}

As their name suggests, dyadic models offer a statistical framework centred on node pairs.  In the Erd\"{o}s-R\'{e}nyi model on $|\mathcal{N}_g|=N$ vertices, the class of probability distributions over possible networks, $\mathcal{P}$, is indexed by the probability $p \in (0,1)$ that a link is formed (independently) between any two nodes.  \citeasnoun{zsg2006}, for example, used a heterogeneous version of this simple random graph model to obtain estimates for the total size of hard-to-count populations.   Observed heterogeneity has also been incorporated via dyadic models that expand on this model just as a probit or logit model generalise a simple Bernoulli statistical model:

$$W_{ij} = \mathbf{1}(X_{ij}^\top \beta + \epsilon_{ij} \ge 0),$$

\noindent which can be used on directed or undirected settings.  The covariates, $X_{ij}$, involve individual-specific variables as well as variables defined for the pair (like geographic distance) or aggregating individual variables into pair-specific quantities.\footnote{In a recent article, \citeasnoun{comolafafchamps2017} extend the model to accommodate misreporting/discordance of elicited links by the individuals involved, which is not uncommon in survey-based network data.}   For example, the metric $f(X_i,X_j)=|X_i - X_j|$ between sender and receiver is usually of interest and gathers how relevant similarity (or homophily) in those variables is for link formation.  (The coefficient on $X_{ij}$ is sometimes called the ``homophily parameter'' by certain authors in the statistics literature.)  Other pairwise functions possibly used include the inner product or projection functions, though others can also be devised.  Whereas estimates can be obtained by usual methods (e.g., likelihood, method of moments), inference in this context typically pays special attention to potential correlations across pairs.  If there are individual factors that are common to all observations related to a given node, dyads will not be independent and $\mathbb{E}(\epsilon_{ij}\epsilon_{kl})$ is potentially nonzero when $i=k, j=l, i=l$ or $j=k$.  In a linear regression model, \citeasnoun{fafchampsgubert2007}, for example, suggest the following robust variance estimator:

$$ \frac{1}{N-K} (\mathbf{X}^\top \mathbf{X})^{-1} \left( \sum_{i,j,k,l} \frac{m_{ijkl}}{2N} X_{ij} \hat \epsilon_{ij} \hat \epsilon_{kl} X_{kl}^\top \right) (\mathbf{X}^\top \mathbf{X})^{-1},$$

\noindent where the matrix $\mathbf{X}$ stacks the dyadic covariates and $m_{ijkl}$ is one when $i=k, j=l, i=l$ or $j=k$ and zero, otherwise.  They also discuss standard errors for the logit (see their footnote 7) and additional examination is provided in \citeasnoun{cameronmiller2014}.\footnote{Related works, focussing on linear models, are \citeasnoun{aronowetal2015} and \citeasnoun{tabordmeehan2019}.}  

It is possible to extend the  Erd\"{o}s-R\'{e}nyi model to incorporate other features.  Consider, for example, the well-known $p_1$ dyadic model offered by \citeasnoun{hollandleinhardt1981} for directed networks.  Their model, which also belongs to the exponential family, postulates that:
\begin{eqnarray} \label{eq:hl1}
\mathbb{P}(W_{ij}=W_{ji}=1) \propto \exp(\alpha^{\text{rec}} + 2\alpha + \alpha^{\textrm{out}}_i + \alpha^{\textrm{in}}_i + \alpha^{\textrm{out}}_j + \alpha^{\textrm{in}}_j)\end{eqnarray}
and
\begin{eqnarray} \label{eq:hl2}
\mathbb{P}(W_{ij}=1, W_{ji}=0) \propto \exp(\alpha + \alpha^{\textrm{out}}_i + \alpha^{\textrm{in}}_j).
\end{eqnarray}
Here, the parameter $\alpha^{\textrm{out}}_i$ encodes the tendency of node $i$ to send out links irrespective of the target (its ``gregariousness''), and $\alpha^{\textrm{in}}_j$ captures node $j$'s tendency to receive links regardless of the sender's identity (its ``attractiveness'').  Since they regulate individual link probabilities, these parameters drive heterogeneity in individual (in and out) degrees across agents in the group.  They also influence how dense or sparse the network tends to be: the more negative those parameters are, the less likely a link is.  The parameter $\alpha^{\text{rec}}$ registers the tendency for directed links to be reciprocated: large, positive values of $\alpha^{\text{rec}}$ will increase the likelihood of symmetric adjacency matrices.     (Note that when $\alpha^{\textrm{out}}_i=\alpha^{\textrm{in}}_i=\alpha^{\text{rec}}=0$ for all $i$, links form independently with probability given by $\exp(\alpha)/(1+\exp(\alpha))$, and the model would correspond to a logit.)  Given an observed network, the authors suggest estimating the model above by Maximum Likelihood and offer an iterative scaling algorithm to ease computation, given the possibly large number of parameters to be estimated.  

This model has been generalised and expanded upon.  \citeasnoun{hoff2005}, for example, considers an augmented model where \emph{multiplicative} interactions between individual unobserved factors are added to the probability specification above (i.e., $\mathbf{z}_i \times \mathbf{z}_j$, where $\mathbf{z}_i$ is a vector of $i$-specific factors), and those plus the additive ``gregariousness'' and ``attractiveness'' features defined previously (i.e., $\alpha_i^{\textrm{out}}$ and $\alpha_i^{\textrm{in}}$) are modeled as random effects.  It is also possible to add covariates, as done in the above mentioned article by \citeasnoun{hoff2005}, who estimates his model using Bayesian methods.  

\citeasnoun{dzemski2018} focusses on such a model, where a link from node $i$ to node $j$ obeys:
$$W_{ij} = \mathbf{1}(X_{ij}^\top \beta + \alpha^{\textrm{out}}_i + \alpha^{\textrm{in}}_j + \epsilon_{ij} \ge 0),$$
where $\epsilon_{ij}$ is a standard normal random variable.  As in \citeasnoun{hollandleinhardt1981}, the parameters $\alpha^{\textrm{out}}_i$ and $\alpha^{\textrm{in}}_j$ appear as fixed effects, hence allowing for an arbitrary correlation between those and with any observed characteristic in the model.  Reciprocity in link formation is captured by a non-zero correlation $\rho$ between $\epsilon_{ij}$ and $\epsilon_{ji}$, which are jointly normal.\footnote{One alternative is to follow \citeasnoun{hollandleinhardt1981} and encode reciprocity in a parameter $\alpha^{\text{rec}}$.   This would lead to a model akin to $W_{ij} = \mathbf{1}(X_{ij}^\top \beta + \alpha^{\textrm{out}}_i + \alpha^{\textrm{in}}_j + \alpha^{\text{rec}} W_{ji} + \epsilon_{ij} \ge 0)$.  This model is nonetheless compatible with multiple joint distributions for $(W_{ij}, W_{ji})$ and, thus, ``incomplete.'' In this case, the model is akin to a game between $i$ and $j$ and point-identification is not guaranteed (see \citeasnoun{depaula2013}).  This is not the case in the original model proposed by \citeasnoun{hollandleinhardt1981} (see displays (\ref{eq:hl1}) and (\ref{eq:hl2})), as the probability distribution over $(W_{ij}, W_{ji})$ is well-defined there. \label{ft:multiplicity}}  Otherwise, errors are independent across pairs.  As pointed out in \citeasnoun{graham2017} (in an undirected network framework), the presence of individual effects encoded in $\alpha^{\textrm{out}}_i$ and $\alpha^{\textrm{in}}_j$ may obscure the identification and estimation of the ``homophily parameter'' $\beta$ as a gregarious individual (i.e., high $\alpha^{\textrm{out}}_i$) will tend to send links indiscriminately, even if there is a homophilous tendency (i.e., high, negative $\beta$ on the distance between her and another node on a particular feature $X$).  It is important thus to control for those, should degree heterogeneity reflect such sources of heterogeneity.

Because each of the $N$ individuals has at her disposal $N-1$ potential liaisons, the setting is akin to a panel where both dimensions (``$N$'' and ``$T$'') are roughly comparable.  Dzemski thus uses tools from the (large-$N, T$) panel data literature (see \citeasnoun{fernandezvalweidner2016} and \citeasnoun{fernandezvalweidner2018}) to establish large sample properties for the estimator.   When the network is sufficiently dense (i.e., the probability of link formation between any two pairs does not vanish asymptotically), the Maximum Likelihood estimator for $\beta$ is such that
$$\hat \beta \quad \stackrel{a}{\sim} \quad \mathcal{N}(\beta_0+B_N, V_N),$$
where the asymptotic bias $B_N$ disappears for large $N$ and the variance decreases with $N$.\footnote{The estimate for $\rho$ is obtained in a second step once the ML estimate for $\beta$ is available.}  (Since there are $N(N-1)$ ordered pairs, the convergence rate is the usual parametric rate.)  Because the number of individual parameters is proportional to $N$, an incidental parameters problem yields an asymptotic bias (see \citeasnoun{fernandezvalweidner2018}).    While the estimator is consistent, the asymptotic bias matters for inference.
 He provides a test of the model based on the prevalence of transitive triads (i.e., vertex triples where links are transitive), which are not used in estimating it, and an application to the microfinance-related networks collected and analyzed in \citeasnoun{bcdj2014}.\footnote{Interestingly, the estimated distribution of ``gregariousness'' and ``attractiveness'' appear to cluster in a few groups, suggesting group-level heterogeneity.}  

\citeasnoun{yanetal2019} provide a related analysis for a similar framework, where the errors $\epsilon_{ij}$ are independent (and $\rho$ is thus zero) logistic random variables.  Under these conditions, the likelihood function for the model is given by
\begin{eqnarray} \label{eq:likeli}
\exp \left(\sum_{i,j} w_{ij} X_{ij}^\top \beta + \alpha^{\textrm{in} \top} \mathbf{d}^{\textrm{in}} + \alpha^{\textrm{out} \top} \mathbf{d}^{\textrm{out}}-\mathcal{C}(\beta, \alpha^{\textrm{in}}, \alpha^{\textrm{out}}) \right),
\end{eqnarray} 
where $\mathcal{C}(\beta, \alpha^{\textrm{in}}, \alpha^{\textrm{out}}) \equiv \sum_{i \ne j} \ln \left( 1+\exp\left( X_{ij}^\top \beta + \alpha^{\textrm{out}}_i + \alpha^{\textrm{in}}_j \right) \right)$ is a normalising constant and $\alpha^{\textrm{in}}, \alpha^{\textrm{out}}, \mathbf{d}^{\textrm{in}}$ and $\mathbf{d}^{\textrm{out}}$ are $N$ dimensional vectors stacking the $\alpha^{\textrm{in}}_i, \alpha^{\textrm{out}}_i$ parameters, and in-degree and out-degree sequences, respectively.  They also rely on similar (though not identical) large $N, T$ panel data manipulations to handle the incidental parameters bias in estimating $\beta$, but pay special attention to the estimation of the individual specific parameters $\alpha$.  Their setup relaxes some of the (non-)sparsity constraints in \citeasnoun{dzemski2018}, which places a lower bound on the likelihood of link formation (see above), allowing the $\alpha$ parameters to drift in value at a $\ln N$ rate.  Like that paper, they also demonstrate the performance of the estimator in simulations and with empirical applications.

In sparser settings, where the errors $\epsilon_{ij}$ are independent logistic random variables, \citeasnoun{charbonneau2017} (and, for undirected networks, \citeasnoun{graham2017}\footnote{This article was previously circulated as ``An Empirical Model of Network Formation: Detecting Homophily when Agents are Heterogeneous.''}, see discussion below) adapts insights from the nonlinear panel data literature on short (i.e., fixed $T$) horizons going back to \citeasnoun{rasch1960}, to ``difference out'' the individual parameters in this logistic regression (see, e.g., Section 5.1 in \citeasnoun{arellanohonore2001}).  The probability mass function for links is the exponential family model in (\ref{eq:likeli}).  Given the properties of the exponential family, the in-degree and out-degree vectors $\mathbf{d}^{\textrm{in}}$ and $\mathbf{d}^{\textrm{out}}$ are sufficient statistics for the parameter vectors $\alpha^{\textrm{in}}$ and $\alpha^{\textrm{out}}$.  This means that the conditional probability for the edges $W_{ij}$s given those random vectors does not depend on the incidental parameters $\alpha^{\textrm{in}}$ and $\alpha^{\textrm{out}}$, thus providing a (conditional) likelihood function.  Unfortunately, such conditional likelihood is computationally complex.  It is nonetheless still possible to condition the incidental parameters out.  Consider, for example, four generic vertices $i,j,l$ and $k$.  The idea is to focus on the probability for one of the two configurations, A or B, in Figure 1 below conditional on either one of the two patterns occurring.  

%\begin{center}
%FIGURE 1 HERE
%\end{center}

\begin{figure}[hbpt]
\begin{center}
\includegraphics[scale = 0.9]{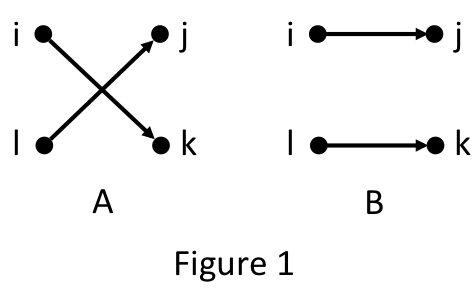}
\end{center}
\end{figure}

It is not hard to see that the conditional probability that $l$ sends a link to $j$ (i.e., $\{W_{lj}=1\}$) as in configuration A, given that only one of those links is in place (i.e., $\{ W_{lj}+W_{lk}=1 \}$) equals:
$$\mathbb{P}(W_{lj}=1 | \mathbf{X}, \alpha, W_{lj}+W_{lk}=1) = \frac{\exp[(X_{lj}-X_{lk})^\top \beta + \alpha^{\textrm{in}}_j - \alpha^{\textrm{in}}_k]}{1+\exp[(X_{lj}-X_{lk})^\top \beta + \alpha^{\textrm{in}}_j - \alpha^{\textrm{in}}_k]},$$
where $\mathbf{X}$ stacks the covariates to all the pairs involved.  Note that the expression above does not depend on $\alpha^{\textrm{out}}_k$.  Likewise, the likelihood that $i$ sends a link to $j$ (i.e., $\{W_{ij}=1\}$), as in configuration B, given that only one link is sent to either one of these potential counterparts (i.e., $\{ W_{ij}+W_{ik}=1 \}$) is
$$\mathbb{P}(W_{ij}=1 | \mathbf{X}, \alpha, W_{ij}+W_{ik}=1) = \frac{\exp[(X_{ij}-X_{ik})^\top \beta + \alpha^{\textrm{in}}_j - \alpha^{\textrm{in}}_k]}{1+\exp[(X_{lj}-X_{lk})^\top \beta + \alpha^{\textrm{in}}_j - \alpha^{\textrm{in}}_k]},$$
which again does not depend on $\alpha^{\textrm{out}}_i$.  These equations look like a conventional panel logit model with regressors $X_{\cdot j}-X_{\cdot k}$ and $ \alpha^{\textrm{in}}_j - \alpha^{\textrm{in}}_k$ as a fixed effect.  A similar manipulation can then be carried out focussing on the event that $i$ sends a link to either $k$ or $j$ (but not both) (i.e., $\{ W_{ij}+W_{ik}=1 \}$), $l$ sends a link to either $k$ or $j$ (but not both) (i.e., $\{ W_{lj}+W_{lk}=1 \}$) and $i$ and $l$ do not send links to the same counterpart (i.e., $\{ W_{lj}+W_{ij}=1 \}$).  Given this event, the probability that $l$ sends a link to $j$ can be seen to be:
\begin{eqnarray*}
\mathbb{P}(W_{lj}=1 | \mathbf{X}, \alpha^{\textrm{in}}, \alpha^{\textrm{out}}, W_{lj}+W_{lk}=1, W_{ij}+W_{ik}=1, W_{lj}+W_{ij}=1) = & & \\
\frac{\exp[((X_{lj}-X_{lk})-(X_{ij}-X_{ik}))^\top \beta]}{1+\exp[((X_{lj}-X_{lk})-(X_{ij}-X_{ik}))^\top \beta]},
\end{eqnarray*}
where the individual parameters are no longer present!  This strategy offers a (quasi-)likelihood function that can be used to estimate of $\beta$.  Since it does not depend on the incidental parameters, it is not restricted by the constraints on sparsity that the previous results rely on, but uses fewer data points.  \citeasnoun{jochmans2018} obtains the large sample properties for the estimator thus defined.  Under regularity assumptions, he shows that $\hat \beta$ is asymptotically normal (without the bias term) and $\| \hat \beta - \beta_0 \| = O_p([N(N-1)p_N]^{-1/2})$, where $p_N$ is the proportion of quadruples like those in Figure 1 that contribute to the estimation objective function.  Intuitively, the more of those there are, the more precise the estimator is.  He offers simulations where the performance of the (conditional) maximum likelihood estimator defined above is shown to perform well even for sparser networks.

\citeasnoun{graham2017} investigates a similar model, also with observed covariates and logistic idiosyncratic errors, but in an undirected network:
\begin{equation} \label{eq:graham}
W_{ij} = W_{ji} = \mathbf{1}(X_{ij}^\top \beta + \alpha_i + \alpha_j + \epsilon_{ij} \ge 0).
\end{equation}
There, the links are established with probability proportional to $\exp(\beta^\top X_{ij} +\alpha_i+\alpha_j)$ and can be interpreted as the outcome from a pairwise stable arrangement with transfers (see discussion below).  Here, the distinction between ``sender productivity'' and ``receiver attractiveness'' for a given node is moot since the link is not directional, but the individual parameters $\alpha_i$ can be seen as the propensity by node $i$ to establish connections.  As in the articles discussed above, they are also treated as fixed effects.  In the absence of covariates, this model was christened the $\beta$-model by \citeasnoun{cds2011}, who analyse its features.  \citeasnoun{yanxu2013} establish the large sample properties for its maximum likelihood estimator (for sufficiently dense networks).  In the presence of covariates, \citeasnoun{graham2017} analyses the large sample properties for the maximum likelihood estimator for $\beta$ (and $\alpha$) as in the articles above, but also offers a conditional maximum likelihood estimator for $\beta$ constructed using sufficient statistics for $\alpha$, allowing him to ``condition those parameters out'' and circumvent the incidental parameters problem estimating $\beta$.   Similarly to the directed case, the degree sequence here is such a sufficient statistic.\footnote{This property is also used by \citeasnoun{pelicangraham2019} to test for the presence of externalities in the model, i.e., whether a link between two nodes depends on edges between those nodes and other vertices, conditional on covariates and individual effects.}   As is the case there, the enumeration of possible degree sequence becomes computationally complex even at moderately sized networks and Graham also relies on tetrads to offer a conditional maximum likelihood estimator, the ``tetrad logit,'' that does not depend on the incidental parameters.   The tetrad logit estimator is asymptotically normal as $N$ grows and able to tackle sparser models, albeit at a slower convergence rate.  While for dense models (where network density converges to a constant as $N$ grows), the convergence rate is parametric ($\| \hat \beta - \beta_0 \| = O_p([N(N-1)/2)]^{-1/2}$), it is $O_p([N(N-1)/2]^{-1/4})$ for sparse network sequences where the density converges to zero at a rate proportional to $N$.

Recent efforts have focussed on extending the models above in several directions.  \citeasnoun{toth2018} and \citeasnoun{gao2018}, for example, study semiparametric versions of the undirected network model (\ref{eq:graham}) without a distributional assumption on $\epsilon_{ij}$.  \citeasnoun{toth2018} examines the identification and estimation of this variation of the model adapting ideas around \citeasnoun{han1987}'s (semiparametric) maximum rank correlation estimator.  \citeasnoun{gao2018} studies the semiparametric identification of the model, but also discusses non-separable models where the index $X_{ij}^\top \beta + \alpha_i + \alpha_j + \epsilon_{ij}$ is replaced by more general forms like $\phi(X_{ij}, \alpha_i, \alpha_j) + \epsilon_{ij}$.   Other extensions veer a bit farther from model (\ref{eq:graham}): \citeasnoun{shichen2016}, for example, examine the estimation of a ``double hurdle'' model\footnote{Traditionally, multiple hurdle models refer to variations of \citeasnoun{cragg1971}.  In the binary outcome context, the model is related to the ``partially observable'' model presented in \citeasnoun{poirier1980}} where
\begin{equation*} 
W_{ij} = W_{ji} = \mathbf{1}(X_{ij}^\top \beta + \alpha_i + \epsilon_{ij} \ge 0) \times \mathbf{1}(X_{ji}^\top \beta + \alpha_j + \epsilon_{ji} \ge 0).
\end{equation*}
The condition above can be interpreted as a pairwise stability requirement on a model where individual utility from links depends additively on every direct connection and there are no externalities (see discussion below).  Here the link is observed whenever it is beneficial to both parties involved, and not if it is detrimental to at least one of them.  A related model (without individual effects) is considered by \citeasnoun{comolafafchamps2014} to study whether (directionally) elicited links are unilaterally or bilaterally formed.

\section{Beyond Dyads}

While dyadic models provide an important angle on the study of link formation, it is plausible that a connection between two nodes depends on liaisons with other nodes in the group.  In a directed network, one such specification would be:
\begin{equation} \label{eq:externalities}
W_{ij} = \mathbf{1} \left( \gamma \sum_{k \ne j} W_{jk} + \epsilon_{ij} \ge 0 \right),
\end{equation}
where for simplicity I abstract from covariates $X_{ij}$ and individual fixed effects $\alpha$.  The individual link formation may also depend on the remaining connections in the network in more general ways.  Here, node $i$ sends an edge to $j$ taking into account those to whom $j$ links to (including whether $j$ is connected to $i$).   One relevant feature of the model above is the econometric endogeneity issue from having $\sum_{k \ne j} W_{jk}$ as a covariate.  If each of the elements in this sum is simultaneously determined according to a expression like (\ref{eq:externalities}), this covariate will be related to $\epsilon_{ij}$.  This would require caution in using ``partial information'' methods (focussed on dyadic likelihoods, for instance) to estimate the parameters above.  It might still nonetheless allow for the use of ``full information'' methods, if a joint distribution for all the links is consistent with the conditionally specified model in (\ref{eq:externalities}).  Importantly though, such models are well known to be ``incomplete'':  they may not produce a unique joint distribution for $W_{ij}, i,j=1,\dots,N$ (see footnote \ref{ft:multiplicity}).  If there are only two nodes, for example, $W_{12} = \mathbf{1} \left( \gamma W_{21} + \epsilon_{12} \ge 0 \right)$ and $W_{21} = \mathbf{1} \left( \gamma W_{12} + \epsilon_{21} \ge 0 \right)$.  If $\gamma$ is positive so that there is a tendency to reciprocate links, there are realisations of $\epsilon_{12}$ and $\epsilon_{21}$ for which the model predicts both $(W_{12}, W_{21})=(0,0)$ and $(W_{12}, W_{21})=(1,1)$ as solutions to this system of equations.  In the statistical literature, this corresponds to the conditionally specified model being ``incompatible'' (see \citeasnoun{arnoldpress1989}).  In econometrics, such models are referred to as ``incomplete'', lacking a selection protocol between $(W_{12}, W_{21})=(0,0)$ and $(W_{12}, W_{21})=(1,1)$ in the example above, for example, and commonly found in the econometric study of strategic interactions with discrete actions (see \citeasnoun{depaula2013}).

A modelling alternative that avoids these issues relies on temporal restrictions and postulates, for example, that
$$
W_{ij, t} = \mathbf{1} \left( \rho W_{ij, t-1} + \gamma \sum_{k \ne j} W_{jk, t-1} + \epsilon_{ij, t} \ge 0 \right).
$$
Here, the edge $W_{ij, t}$ between $i$ and $j$ in period $t$ depends on lagged network features $W_{ij, t-1}$ and $\sum_{k \ne j} W_{jk, t-1}$, which are predetermined with respect to the innovation $\epsilon_{ij, t}$.   When empirically adequate, this specification circumvents the simultaneity issue and allows for the addition of other covariates $X_{ij,t}$, while requiring repeated observations on the network.\footnote{See also the related discussion of temporal lags in social interactions models and their identification by \citeasnoun{manski1993}, Section 4.  \citeasnoun{bdf2020} provide a review of that literature in this volume.}  It is also possible to include individual parameters $\alpha$, though proper attention should be given to the dynamic panel data nature of the setting in this case (see, e.g., \citeasnoun{arellanohonore2001}, Section 8, for the conventional panel data treatment).  Such a model (with individual effects) is indeed contemplated in \citeasnoun{graham2016}, where similar conditioning arguments as those in \citeasnoun{charbonneau2017} or \citeasnoun{graham2017}, though conditional on different events, are used to examine (semiparametric) identification and estimation on undirected networks.\footnote{See also \citeasnoun{hanhsiehko2019}.}

Another strategy to bypass the simultaneity issues referred to above is to model not only dyads, but also triads, tetrads and more general $k$-tuples directly.  \citeasnoun{chandrasekharjackson2016} propose one such framework, which they call subgraph generation model (SUGM).  It specifies a set of $K$ subgraph classes, potentially involving more than two nodes each and probabilities for each of these two nodes.  If one has $K=2$, for example, one of those subgraph classes could be taken to be the possible graphs between two nodes and the second, the possible networks among three vertices.  One then forms subgraphs at random from each of these classes among all the tuples in the group and the final network is obtained by taking the union of all the edges thus sampled.  Some of these edges may be redundant as a link between two nodes could have appeared both as a draw from the two-node subgraph class and the three-node one.  The subgraphs may also get ``meshed'' together: a complete subnetwork among nodes $i, j$ and $k$ could have arisen as a genuine such subgraph on that triad or three independent complete graphs on the three pairs involving those three vertices.  Disentagling the count of subgraphs in the model that are genuinely formed or just happenstance from the composition of other subgraphs can be done by noting that the counts of possible subgraphs is a mixture of both genuinely and incidentally formed subgraphs.  This renders a system of equations that can then be solved for the parameters of interest and potentially matched to moments for the desired parameters.

\subsection{Strategic Formation Models}

An alternative is to directly handle the challenges highlighted above.  Since econometric models with interacting agents are natural platforms where such difficulties arise, much of the work speaking to such issues corresponds to estimable game theoretic frameworks.   (While the economic narrative is not as salient in the statistical models described previously, many of those can also be interpreted through the lens of behavioural models.) 

The first step in framing the network formation as a game involves a specification for the set of players, their actions and payoffs.  In the present setting, the players are represented by the group of nodes $\mathcal{N}_g$ to be eventually connected in equilibrium, corresponding for example to individuals, households or firms.  Their actions are in turn related to the formation of links.  When the model is aimed at a directed network characterisation, actions encode whether a node sends a link to another node.  In undirected environments, on the other hand, the connections might need to be agreed upon.  For a given graph $g \equiv (\mathcal{N}_g, \mathcal{E}_g$) designating how players are connected, a node $i$ is assigned a network dependent payoff.  Whereas the payoff specification will depend on the context, a common parameterisation (on an undirected graph, see e.g. \citeasnoun{de_Paula_et_al_ECMA2018}) is given by:
\begin{equation} \label{eq:fof}
 U_i(g) \equiv \sum_{j \ne i} W_{ij} \times \big( u + \epsilon_{ij} \big) + 
\left| \cup_{j:W_{ij}=1} N_j(g) - N_i(g) - \{ i \} \right| \nu +  \sum_{j} \sum_{k>j} W_{ij} W_{ik} W_{jk} \omega,
\end{equation}
where  $W_{ij}=1$ if $i$ and $j$ are connected, $N_i(g)$ denotes the set of nodes directly connected to node $i$ in the graph and $|\cdot|$ is the cardinality of a given set.   The first term on the right-hand side of (\ref{eq:fof}) registers the payoff from direct connections and involves the parameter $u$ and idiosyncractic variables $\epsilon_{ij}$, unobserved by the econometrician.  The second term encodes the utility obtained from indirect connections: $\left| \cup_{j:W_{ij}=1} N_j(g) - N_i(g) - \{ i \} \right|$ is the number of individuals connected to direct couterparties of $i$, but not directly connected to her.  Finally, the last term summarises any benefits accruing from two direct connections also being connected and induces incentives for ``clustering,'' a commonly observed phenomenon.   While I omit observable covariates in the expression above, the parameters ($u, \nu, \omega$) can also be made to depend on those.  

It is also important to clearly establish the information structure.  In the literature so far (and in accordance with the statistical models discussed above), it is often assumed that information is complete so agents are informed about others' (observed and unobserved) payoffs and incentives perfectly.  Incomplete information, albeit possibly more plausible and epistemically more adequate in certain contexts, has less often been analysed in the literature.

Another feature of the environment that requires attention relates to transferability.  This refers to the possibillity for agents to transfer payoffs among themselves, not always monetarily but also possibly through other means.  When available, this possibility allows nodes, for example, to bid for their preferred counterparties by accepting lower payoffs themselves.  At two opposite ends of the spectrum are non-transferable utility (NTU) models, when there is no technology enabling agents to decrease their utility to benefit a potential partner, and transferable utility (TU) models, which allows transfers of utility at a constant ``exchange rate'' and the total gain from the matching (surplus) is what matters for stability of the relationship.\footnote{Another possibility is the intermediate scenario with imperfect transferable utility (ITU) where transfers are allowed, but at an ``exchange rate'' between individual utilities that is not constant and possibly endogenous to the economic environment.  While this would also be categorised as transferable utility, the conventional terminology focusses on the constant ``exchange rate'' case (see \citeasnoun{Chiappori_Book2019}).}  Which one is adopted again depends on the context at hand.

In closing the model, one then relies on a solution concept prescribing how individual behaviours are aggregated to generate an equilibrium network.   Whereas traditional concepts in game theory (e.g., Nash equilibrium) can be envisioned and adapted to this case, the theoretical literature has offered additional notions to better capture the peculiarities of certain network formation contexts.   When modelling undirected networks, for example, \citeasnoun{Jackson_Wolinsky_JET96} propose pairwise stability as an alternative solution concept.  A network $g$ is pairwise stable according to \citeasnoun{Jackson_Wolinsky_JET96} if any link present in $g$ is mutually beneficial and any absent link is detrimental to at least one of the parties involved.  More formally,
$$\forall ij \in g, \ U_i(g) \geq U_i(g_{-ij}) \text{ and } U_j(g) \geq U_j(g_{-ij})$$
and
$$\forall ij \notin g, \  U_i(g) > U_i(g_{+ij}) \text{ or } U_j(g) > U_j(g_{+ij}),$$
where $ij \in(\notin)~g$ signifies that the link between $i$ and $j$ pertains (not) to the set of edges in $g$.\footnote{This is related, but different from the stability concept typically used in ``marriage market'' models.}  The network $g_{-ij}$ is $g$ without the link beween $i$ and $j$ and $g_{+ij}$ is the network $g$ with the link beween $i$ and $j$.   The theoretical literature has contemplated several variations to this stability concept (e.g., pairwise Nash stability, strong stability, see the discussion in \citeasnoun{jackson2009}) and \citeasnoun{BlochJackson_IJGT2006} adapt this solution concept to an NTU environment.

For a given parameter vector and observable covariates, the model translates the distribution of unobservable variables (e.g., $\epsilon_{ij}$) into a probability distribution over the equilibrium sets for the game described above.  Since this probability distribution is indexed by the parameter vector, this delivers a statistical model on which one can in principle perform estimation and inference.  One difficulty in doing this is that there might be more than one stable network for given parameter value and realisations of observable and unobservable random variables.  This is related to the statistical difficulties highlighted earlier in this section and is potentially problematic for identification (i.e., the reverse mapping between observed distributions and parameters), computation and inference.

Another possibility is to rely on iterative procedures where the presence or absence of links is evaluated as individuals or pairs take turns in a random meeting protocol as in the ``preferential attachment'' framework mentioned earlier, where agents form links sequentially and the establishment of new links is more likely with higher-degree existing nodes (see \citeasnoun{barabasialbert1999}).\footnote{It pays to note that such stochastic revision processes are not unrelated to the (non-iterative) equilibrium notions mentioned earlier and discussed below.  For example, in a non-transferable utility setting, \citeasnoun{Jackson_Watts_JET02} demonstrate that a process where pairs meet sequentially and are offered to (myopically) form or maintain links that are mutuallly beneficial and dissolve links that are not beneficial to at least one of the parties involved converges to a pairwise stable network or a cycle.}    Before discussing models relying on the equilibrium notions above, I will first discuss alternatives relying on iterative protocols.

\subsubsection{Iterative Network Formation}

\citeasnoun{Mele_ECMA2017}, \citeasnoun{Christakis_et_al_NBER10} and \citeasnoun{Badev_WP18} are notable examples of strategic network formation models based on a stochastic meeting protocol whereby individuals or pairs sequentially revise their links.  While such protocols are not meant to be directly fit to the data, the random meeting sequences and unobservable errors guiding the decisions to establish or interrupt connections lead to a potentially estimable distribution over networks.  \citeasnoun{Mele_ECMA2017} studies a directed network, and  \citeasnoun{Badev_WP18} expands the analysis there to the joint determination of links and behaviours.\footnote{\citeasnoun{Mele_ECMA2017} was previously circulated as ``A Structural Model of Segregation in Social Networks'' (2015).  Previous versions for \citeasnoun{Badev_WP18} appeared as ``Discrete Games with Endogenous Networks: Theory and Policy'' (2013) and "Discrete Games in Endogenous Networks: Equilibria and Policy" (2017).}  \citeasnoun{Christakis_et_al_NBER10}, on the other hand, examine an undirected network.  The last two present empirical applications to links among adolescents using the AddHealth data and \citeasnoun{Mele_AEJ2019} applies the methodology in \citeasnoun{Mele_ECMA2017} to examine segregation in high schools, also using AddHealth data.

\citeasnoun{Mele_ECMA2017} models a directed network where $W_{ij}=1$ if individual $i$ offers a link to individual $j$, and $=0$, otherwise.  The utility function for individual $i$ is given by:
 $$U_i(g) \equiv \sum_{j \ne i} W_{ij} u_{ij}^\theta + \sum_{j \ne i} W_{ij} W_{ji} m_{ij}^\theta + \sum_{j \ne i} W_{ij} \sum_{k \ne i,j} W_{jk} \nu_{ik}^\theta +  \sum_{j \ne i} W_{ij} \sum_{k \ne i,j} W_{ki} \nu_{kj}^\theta,$$
where $u_{ij}^\theta \equiv u(X_i,X_j;\theta)$ represents the utility from directly linking to other individuals and the first term involving $\nu_{ij}^\theta \equiv \nu(X_i,X_j;\theta)$ encodes the utility from indirectly linking to a friend's friend.  These then play the same role as $u$ and $\nu$ in the utility function previously introduced.  Since this is a directed network, $m_{ij}^\theta \equiv m(X_i,X_j;\theta)$ marks the utility from a mutual, reciprocated link.  (In undirected networks, links are reciprocated by definition and this term does not show up in the utility function presented earlier in this review.)  The second term involving $\nu_{ij}^\theta$ internalises some of the impact a link to individual $j$ generates for individuals that had offered links to $i$.\footnote{The author refers to this as popularity: ``When an agent forms a link, he/she automatically creates an indirect link for other agents that are connected to him/her, thus generating externalities and impacting his/her `popularity.'''}  Utility is non-transferable and information is complete.

There is then a meeting sequence $m = \{ m^t \}_{t=1}^\infty$, where $m^t = (i,j)$ means that $i$ can offer or dissolve a link to $j$ in iteration $t$.  Let $\mathbb{P}(m^t=ij | g^{t-1},X) = \rho(g^{t-1},X_i,X_j),$ where $g^{t-1}$ is the network in iteration $t-1$.  In addition, the meeting probability between $i$ and $j$ does not depend on there being a link between them and each meeting has positive probability of occuring (i.e., $\rho(g^{t-1},X_i,X_j)=\rho(g^{t-1}_{-ij},X_i,X_j)>0, \forall ij$). This ensures that the likelihood function from this model does not depend on the meeting protocol.

Finally, whenever $i$ is offered a meeting with another node, it is supposed that she receives idiosyncratic shocks $(\epsilon_1,\epsilon_0)$ to the utility of forming a link ($W_{ij}=1$) or not ($W_{ij}=0$).  An edge to $j$ is established if and only if
$$U_i(W^t_{ij} = 1, g^t_{-ij}, X; \theta) + \epsilon_{1t} \ge U_i(W^t_{ij} = 0, g^t_{-ij}, X; \theta) + \epsilon_{0t}.$$
\noindent While there were no unobservable errors up to this point, once given the choice the decision by $i$ acts much like a standard random utility model over the remaining $|\mathcal{N}_g|-1$ nodes.  (This implies that the average degree is roughly proportional to $|\mathcal{N}_g|$ and the model produces a dense network.)  This revision process then leads to subsequent additions and deletions of edges where the resulting process forms a Markov chain on networks $\{ g^t \}$.  In the absence of unobservable shocks (i.e., if $\epsilon_0$ and $\epsilon_1$ are zero with probability one), the chain converges to one of the Nash equilibria for the game without $\epsilon$s.  Under the assumption that the $\epsilon$s are distributed independently (across time and links) and follow an Extreme Value Type I distribution, it instead converges to a unique stationary distribution:
$$\pi(g,X;\theta) = \exp[Q(g,X;\theta)-\mathcal{C}(\theta)],$$
where $\mathcal{C}(\theta) = \ln\{ \sum_{g'} \exp[Q(g',X;\theta)] \}$ and
$$Q(g,X;\theta) = \sum_{(i,j)} W_{ij} u_{ij}^\theta + \sum_{(i,j)} W_{ij} W_{ji} m_{ij}^\theta + \sum_{(i,j,k)} W_{ij} W_{jk} \nu_{ik}^\theta.$$ 
As noted by the author, the Nash equilibria for the game with payoffs thus defined correspond to the maxima of the function $Q(g,X;\theta)$.

This distribution can in principle be taken to data on either one or more networks.  In fact, the stationary distribution above describes an exponential random graph model, discussed previously in this survey.  The model is estimated by Bayesian methods, producing a posterior distribution over parameters of interest.  One important difficulty with such models is nonetheless the computation of the normalising constant $\mathcal{C}(\theta)$ in $\pi(g,X;\theta)$ as the summation is over all possible directed networks between the individuals in the group.   With 10 individuals, for example, there are $2^{90} \approx 10^{27}$ such network configurations.\footnote{Similar issues also appear in models like those in display (\ref{eq:likeli}), which also belong to the exponential family.}  This is relevant above since it is necessary to compute such denominator to solve for the posterior distribution either analytically or numerically (e.g., via simulations of the posterior distribution).

Alternative strategies to circumvent this issue or to approximate the denominator in other settings include pseudo-likelihood methods (\citeasnoun{besag1975}, \citeasnoun{straussikeda1990}) and variational principles (see \citeasnoun{jordanwainwright2008} and the recent article by \citeasnoun{melezhu2019} for an application in econometrics).  These and other protocols are briefly discussed in \citeasnoun{DePaula2017}. \citeasnoun{Mele_ECMA2017} instead handles it using simulation (Markov chain Monte Carlo, MCMC) methods (see also \citeasnoun{kolaczyk2009} and references therein).  The first challenge is the computation of $\mathcal{C}(\theta)$ for a given parameter value $\theta$.  Here, a (Metropolis-Hastings) algorithm to simulate the distribution of networks without the normalising constant can be designed.

Unfortunately, this simulation protocol is itself not immune to problems.  It is well known, for example, that parameter changes in ERGMs may lead to abrupt changes in probable graphs (see \citeasnoun{snijders2002}).  In addition, in parameter regions where the distribution over networks is multimodal, the convergence of the algorithm is impractically slow for MCMC protocols where networks change only locally from iteration to iteration (see \citeasnoun{bhamidietal2011}).  Here, the modes of the distribution will correspond to the maxima of the function $Q$ as highlighted previously, which are in turn related to the Nash equilibria for the game discussed previously.  Hence, while the stationary distribution for the Markov process defined by the meeting protocol is unique, equilibrium multiplicity is also a complicating computational feature here.  To accelerate convergence, the article suggests a simulation algorithm that updates networks at larger steps (see Appendix B.1 to the article).

For parameter regions where the distribution is unimodal on the other hand, \citeasnoun{bhamidietal2011} and \citeasnoun{Chatterjee_Diaconis_AOS13} show that graph draws are indistinguishable from a random network model  with independent link formation (i.e., an Erd\"{o}s-R\'{e}nyi model) or mixture of such models.  \citeasnoun{Mele_ECMA2017} also shows that a similar phenomenon occurs in a simplified version of his model, without covariates and positive utility from mutual links ($m>0$).  This points to an identification issue when estimation is based on a single network in this particular example.\footnote{When multiple networks are used, identification is attained with variation in sufficient statistics across networks: ``If the sufficient statistics are not linearly dependent, then the exponential family is minimal and the likelihood is stricly concave, therefore the mode is unique.'' (\citeasnoun{Mele_ECMA2017})}  The result does not allow for covariates, but the author conjectures that the sign of such externalities (on reciprocated links) will remain relevant.

In the Bayesian estimation, one also needs to approximate the posterior distribution for the parameters of interest.  The ``inner'' simulation of the networks for a given parameter vector discussed above is a component of the ``outer'' simulation protocol for the posterior distribution presented in \citeasnoun{Mele_ECMA2017}. Here too, the article follows a Markov chain Monte Carlo procedure based on the exchange algorithm proposed by \citeasnoun{Murray_et_al_UAI06} to avoid the computation of the normalising constant.  The model is taken to data in \citeasnoun{Mele_AEJ2019}, where the author studies ethnic segregation in high schools using data from AddHealth.

\subsubsection{Non-Iterative Network Formation}

An alternative strategy takes a non-iterative perspective where network formation is obtained as a simultaneous move game with players as nodes/vertices and the action space as potential links.  For example, \citeasnoun{leung2015a} studies a directed network formation model under incomplete information and Bayes-Nash as the solution concept adapting (two-step) estimation strategies commonly used in the econometrics on such games (see \citeasnoun{depaula2013}).  He employs the model to analyse trust networks in India.\footnote{An ealier working paper taking a similar modelling framework is \citeasnoun{gilleskiezhang2009}, where the authors also examine smoking behaviour among network members.  \citeasnoun{Ridder_Sheng_WP15} and \citeasnoun{candelariaura2018} are recent working papers also studying incomplete information network formation games.}  Also modelling a directed network, \citeasnoun{Gualdani_WP2019} examines a complete information model employing Nash equilibrium as the solution concept and studies board interlocks among firms.  One important challenge in this context relates to the dimensionality of potential networks and the article offers suggestions to reduce the computational burden.  I will here focus on works using pairwise stability (without transferability).

One avenue is to adapt some of the strategies used in the empirical games literature.   To illustrate this, take the simplified directed network formation setting discussed previously where there are two nodes, 1 and 2, and the payoff obtained by $i=1,2$ from a link to $j \ne i$ is given by: $\gamma + \epsilon_{ij}$ if $j$ also offers a link to $i$, and $\epsilon_{12}$, otherwise.  If $i$ does not send a link to $j$, her payoff is zero.  The payoffs for $j$ are analogously defined.   A (Nash) equilibrium for this game leads to an econometric model where $W_{12} = \mathbf{1} \left( \gamma W_{21} + \epsilon_{12} \ge 0 \right)$ and $W_{21} = \mathbf{1} \left( \gamma W_{12} + \epsilon_{21} \ge 0 \right)$.  As discussed earlier, if there is a tendency to reciprocate links (i.e., $\gamma$ is positive ), there are realisations of $\epsilon_{12}$ and $\epsilon_{21}$ for which the model predicts both $(W_{12}, W_{21})=(0,0)$ and $(W_{12}, W_{21})=(1,1)$ as solutions to this system of equations.   While this does not tie down a single probability distribution for $(W_{12},W_{21})$, it nonetheless offers bounds on the probabilities for the possible realisations of this vector.  The probability for the event $\{(W_{12}, W_{21})=(0,0)\}$ is minimized if $(0,0)$ is never selected when $\epsilon_{12}$ and $\epsilon_{21}$ are such that other possible equilibria exist.  It is on the other hand maximised when $(0,0)$ is always chosen when there are other possible solutions for those realisations of $\epsilon_{12}$ and $\epsilon_{21}$.  These quantities thus bound the probability for the event $\{(W_{12}, W_{21})=(0,0)\}$ and those inequalities in turn provide information on the parameters guiding the data generating process.  Parameters that lead to lower probability bounds above or upper probability bounds below the observed frequency in the data are not consistent with the generating process.  At the same time, more than one parameter will typically be consistent with the data rendering the model ``partially identified'' (see \citeasnoun{tamer2010}).\footnote{For additional strategies to handle the multiplicity problem, see \citeasnoun{depaula2013}.}

To illustrate how this translates with an alternative solution concept, consider, for instance, an undirected network formation game among three individuals ($|\mathcal{N}_g|=3$) with non-transferable utility and complete information.  Let payoffs be given by:  
$$U_i(g) \equiv \sum_{j \in 1,\dots,n, j \neq i} \delta^{d(i,j;g)-1} \left( 1 + \epsilon_{ij} \right) - |N_i(g)|,$$  
where $d(i,j;g)$ is the minimum distance between $i$ and $j$ in the graph $g$, $0<\delta<1$, $\epsilon_{ij}$ are unobservable (to the researcher) preference shocks and $|N_i(g)|$ is the number of direct connections of individual $i$ in graph $g$.  Assume that observed networks are pairwise stable.  For $\epsilon_{ij}=\epsilon_{ji}$, $0<\epsilon_{23}<\delta/(1-\delta)$, we can represent the possible pairwise stable equilibria in the space of unobservables as:

\newpage

\begin{figure}[htbp]
\setlength{\unitlength}{0.12in} % selecting unit length
\centering % used for centering Figure
\begin{picture}(40,30) % picture environment with the size (dimensions)
% 32 length units wide, and 15 units high.
\put(9,29){\makebox(0,0){$\epsilon_{12}$}}
\put(39,6){\makebox(0,0){$\epsilon_{13}$}}
\put(0,7){\vector(1,0){40}}
\put(10,0){\vector(0,1){30}}
\put(10,12){\line(1,0){30}}
\put(15,7){\line(0,1){23}}
\put(8,7){\line(-2,1){8}}
\put(4.5,7){\line(0,1){23}}
\put(10,5){\line(1,-2){2.5}}
\put(10,2){\line(1,0){30}}
\put(4.5,7){\line(0,1){0.5}}
\put(25,20){\makebox(0,0){$\{12, 13\}$}}
\put(12.5,21){\makebox(0,0){$\{12, 13\}$}}
\put(12.5,19){\makebox(0,0){$\{12, 23\}$}}
\put(12.5,11){\makebox(0,0){$\{12, 13\}$}}
\put(12.5,9.5){\makebox(0,0){$\{12, 23\}$}}
\put(12.5,8){\makebox(0,0){$\{13, 23\}$}}
\put(25,9){\makebox(0,0){$\{12, 13\} \quad \{13, 23\}$}}
\put(7,20){\makebox(0,0){$\{12, 23\}$}}
\put(2,20){\makebox(0,0){$\{12\}$}}
\put(2,8){\makebox(0,0){$\{12\} \{23\}$}}
\put(5,7.5){\vector(1,1){1.5}}
\put(7.5,9.5){\makebox(0,0){$\{23\}$}}
\put(25,5){\makebox(0,0){$\{13, 23\}$}}
\put(3,3){\makebox(0,0){$\{23\}$}}
\put(10.5,3){\vector(1,0){2}}
\put(14,3){\makebox(0,0){$\{23\}$}}
\put(7.5,12){\makebox(0,0){$\delta/(1-\delta)$}}
\put(8,6){\makebox(0,0){-1}}
\put(9,5){\makebox(0,0){-1}}
\put(7.5,2){\makebox(0,0){$-1-\frac{\epsilon_{23}}{\delta}$}}
\put(4,6){\makebox(0,0){$-1-\frac{\epsilon_{23}}{\delta}$}}
\put(15,6){\makebox(0,0){$\delta/(1-\delta)$}}
\put(10.5,1){\vector(1,0){2}}
\put(15,1){\makebox(0,0){$\{13\} \{23\}$}}
\put(25,1){\makebox(0,0){$\{13\}$}}
\end{picture}
\end{figure} 

The approach above would produce on bounds on $\delta$ corresponding to the probability that a particular network is pairwise stable (though possibly not unique) (upper bound) and the probability that it is the unique pairwise stable network (lower bound).  In the figure above (which does not comprise the whole space for $\epsilon$s), those bounds for the network $\{12,13\}$ would be
$$ \mathbb{P}(\epsilon_{12}, \epsilon_{13} \ge 0) \ge \mathbb{P}(\{12,13\}) \ge \mathbb{P}(\epsilon_{12}, \epsilon_{13} \ge \delta/(1-\delta)),$$
and one could form similar bounds for all (= 8) possible networks (exploring the whole space of unobservables).  Unfortunately, while this might be conceivable when only three individuals are involved, for even a moderate number of nodes, the number of networks one would need to consider as potential equilibria is computationally intractable.  With more than 24 individuals, for instance, there are more potential networks than atoms in the observable universe.

Given this dimensionality issue, one possible avenue to reduce the computational burden is to focus on smaller subnetworks involving subgroups of individuals.  \citeasnoun{Sheng_ECMA2016} develops this strategy in a model  for undirected networks.  She focusses on pairwise stable networks (either with transferable and non-transferable utility, see previous definitions) under complete information and a utility structure given by:

 $$U_i(g) \equiv \sum_{j \ne i} W_{ij} (u_{ij}^\theta + \epsilon_{ij}) + \frac{1}{|\mathcal{N}_g|-2} \sum_{j \ne i} W_{ij} \sum_{k \ne i,j} W_{jk} \nu +  \frac{1}{|\mathcal{N}_g|-2} \sum_{j,k \ne i} W_{ij} W_{ki} W_{jk} \omega.$$

As before, $u_{ij}^\theta \equiv u(X_i,X_j;\theta)$ recrods the direct utility obtained from linking to other individuals in the group, $\nu$ provides benefits to indirect friendships and $\omega$ indicates additional payoff to having common connections also be linked to each other.\footnote{In contrast to the utility specification in the begining of this chapter, the specification above normalises the number of connections by $|\mathcal{N}_g|-2$ and allows for some double counting: if $i, j$ and $k$ are all connected, each accrues benefits from direct connection with two individuals, indirect connections from each of the other two and the fact that the other two are also linked to each other.}  The unobservable random variables $\epsilon$ are independent and follow a known distribution.  (Since there are preference shocks for each potential link, isolated individuals are unlikely when $|\mathcal{N}_g|$ is large.)  The data is presumed to come from a sample of i.i.d. networks.

Pairwise stability (with either transferable utility or not) is determined by the marginal utility of a link ($\Delta U_{ij}(g) = U_i(g) - U_i(g_{-ij})$ when $ij \in g$ and $\Delta U_{ij}(g) = U_i(g_{+ij}) - U_i(g)$ when $ij \notin g$).  Let then $\textrm{PS}(\Delta U(X,\epsilon))$ denote the set of pairwise stable networks for realisations of $X$ and $\epsilon$.   The probability that one observes $g$ is then given by:
\begin{eqnarray*}
 & \mathbb{P}(g | X) = \int_{g \in \textrm{PS}(\Delta U(X,\epsilon)) \wedge |\textrm{PS}(\Delta U(X,\epsilon)|=1} dF(\epsilon) + & \\
 & \int_{g \in \textrm{PS}(\Delta U(X,\epsilon)) \wedge |\textrm{PS}(\Delta U(X,\epsilon)|>1} \lambda(g | \textrm{PS}(\Delta U(X,\epsilon)) dF(\epsilon), &
\end{eqnarray*}
where $\lambda$ is the probability that $g$ is selected for realisations of $\epsilon$ that allow for multiple pairwise stable networks.  One can then emulate our previous discussion on empirical games to generate bounds on the probability and guide the computation of identified parameter sets.  As previously discussed, these bounds are unfortunately computationally intractable for even moderately sized networks.  Instead, \citeasnoun{Sheng_ECMA2016} suggests looking at subgraphs on $A \subset \mathcal{N}_g$ nodes: the network $g_A$ comprising $A$ and all edges in $g$ linking those nodes.  The probability of observing such a subnetwork is then given by:
\begin{eqnarray*}
 & \mathbb{P}(g_A| X) = \int_{g_A \in \textrm{PS}_A(\Delta U(X,\epsilon)) \wedge |\textrm{PS}_A(\Delta U(X,\epsilon)|=1} dF(\epsilon) + & \\
 & \int_{g_A \in \textrm{PS}_A (\Delta U(X,\epsilon)) \wedge |\textrm{PS}_A(\Delta U(X,\epsilon)|>1} \sum_{g_{-A}} \lambda(g_A | \textrm{PS}(\Delta U(X,\epsilon)) dF(\epsilon), &
\end{eqnarray*}
where the addition over $g_{-A}$ sums over the collection of complementary nodes ($\mathcal{N}_g \backslash A$) and edges in $g$ connecting them to each other and to nodes in $A$.  $\textrm{PS}_A (\Delta U(X,\epsilon))$ is the subset of networks in $A$ that are part of a network in $\textrm{PS}(\Delta U(X,\epsilon))$.  These deliver:
$$ \int_{g_A \in \textrm{PS}_A(\Delta U(X,\epsilon)) \wedge |\textrm{PS}_A(\Delta U(X,\epsilon)|=1} dF(\epsilon) \le \mathbb{P}(g_A | X_A) \le \int_{g_A \in \textrm{PS}_A(\Delta U(X,\epsilon))} dF(\epsilon),$$
where the upper bound is the probability that the subnetwork $g_A$ on nodes $A$ pertains to a pairwise stable network and the lower bound is the probability that the \emph{only} subnetwork on nodes $A$ pertaining to a pairwise stable network is $g_A$.

Consider for example the network formation game on three individuals presented earlier in this subsection.  For the subnetwork $\{12\}$ on nodes 1 and 2, the upper bound is given by the probability that $\epsilon_{12}$ is greater than zero minus the probability for the triangular region where only $\{23\}$ is pairwise stable.  In this region, $\{12\}$ always pertains to a pairwise stable network.  The lower bound is given by the probability that $\epsilon_{12}$ is greater than $\delta/(1+\delta)$ and $\epsilon_{13}$ is greater than zero plus the probability that $\epsilon_{13}$ is less than zero minus the subregions where $\{23\}$ is also a pairwise stable network.  In this region, $\{12\}$ is the only network on nodes 1 and 2 that is part of a pairwise stable network is $\{12\}$ (even though there are multiple pairwise stable networks in the region where $\epsilon_{12}$ is greater than $\delta/(1-\delta)$ and $\epsilon_{13}$ is between zero and $\delta/(1-\delta)$.

In the article, Sheng imposes (exchangeability) restrictions (on equilibrium selection and payoff primitives) that guarantee that these bounds are nontrivial even as the number of individuals in the groups gets larger.
Among of other things, such (exchangeability) restrictions also imply a dense network (as in \citeasnoun{Mele_ECMA2017}): the total number of links is $O_p(N^2)$ (see, e.g., \citeasnoun{Orbanz_Roy_IEEE15}).  While the sets above are not sharp, i.e., more informative bounds on the parameters can intuitively be obtained by considering subnetworks on a larger number of nodes, they are potentially computable.\footnote{Previous versions of the article, as the one reviewed in \citeasnoun{DePaula2017}) also considered simpler bounds: $\mathbb{P}(W_A = w_A | X_A)$ $\le$ $\int_{\exists W_{-A} : w_A \in \textrm{PS}(\Delta U_A(W_{-A},X_A,\epsilon_A))} dF(\epsilon_A)$ and $\mathbb{P}(g_A | X_A) \ge \int_{\forall g_{-A} : g_A \in \textrm{PS}(\Delta U_A(g_{-A},X_A,\epsilon_A)) \wedge |\textrm{PS}_A(\Delta U_A(g_{-A},X_A,\epsilon_A)|=1} dF(\epsilon_A)$.  In words, the upper bound for $\mathbb{P}(g_A | X_A)$ is the probability that subnetwork $g_A$ is pairwise stable for some $g_{-A}$ and the lower bound is the probability that, for any $g_{-A}$, only subnetwork $g_{A}$ is pairwise stable.  These bounds do not require pairwise stability on the rest of the network and are thus easier to compute, but are less informative than the ones above and might yield trivial bounds for larger groups.}  The article offers a computational algorithm to perform such computation and indicates a few additional potential simplifications.  For example, when $\nu$ and $\omega$ are nonnegative (which guarantees existence of pairwise stable networks when utility is non-transferable), the game is supermodular and the solution set possesses a maximal and a minimal element.  This can be leveraged to reduce the computational complexities here as done by \citeasnoun{Miyauchi_JOE16} (also in the context of pairwise stable network formation) and other authors in the empirical games literature.  A Monte Carlo study in the paper demonstrates the performance of the algorithm in a transferable utility context with 50 and 100 networks varying in size from 25 individuals to 100.

Another use of subnetworks to circumvent the challenges presented in this setting appears in \citeasnoun{de_Paula_et_al_ECMA2018}.  The article works on a complete information, non-transferable utility model using pairwise stability as the solution concept.  The treatment is tailored to handle large networks and $\mathcal{N}_g$ is an uncountable set with continuum cardinality, taken to be an approximation for a large group of individuals.  The approach described here starts with a ``large'' network.  Related approaches, also focussed on large networks, but taken as limits for finite sequences of networks is given in \citeasnoun{Leung_WP15}, \citeasnoun{Menzel_WP16} and \citeasnoun{Boucher_Mourifie_EJ17}.  To capture sparsity, they restrict payoffs, allowing only for a finite number of links $L$ such that in equilibrium one obtains a bounded degree graph on the continuum (sometimes referred to as a \emph{graphing}).  Utilities are also assumed to depend only on individual characteristics (and not identities) and on indirect connections only up to a finite distance (depth $D$).  This allows one to focus on ``network types'' defined by one's local neighbourhood, whose cardinality is potentially more manageable than that of networks on individual nodes.

To illustrate the strategy, consider a very simple network formation game where individuals can only form one link and their utility depends only on this link.  Here, both $L$ and $D$ are one.  Nodes are characterised by $X$, which takes two values, $B$ or $W$ and there is a continuum of individuals of each type $\mu_B$ and $\mu_W$.  Outcomes are thus given by ordered pairs $(x,y)$, where $x$ is the individual's characteristic and $y$, her connection's characteristic.  Utilities are given by 
$$U_i(g) \equiv u_{xy} + \epsilon_i(y),$$
where again $x$ marks the individual's charateristic and $y$, her counterpart.  If no link is formed, the payoff is normalised to zero.  Hence, there are four parameters ($f_{x,y}, x, y \in \{B, W\}$ and two preference shocks for each individual: $\epsilon_i(B)$ and $\epsilon_i(W)$.\footnote{Additional restrictions are imposed on the preference shocks so that in equilibrium, even in large networks, there are isolated individuals.}

The paper defines \emph{network types} as the local network surrounding an individual in an observed network.  Given the payoff structure, the network type should record payoff-relevant connections.  In this simple individual, the relevant network type is given by the pair $(x,y)$.  In a more general setting a network type is characterised by the individual herself, her direct connections, their direct connections and so on together with each of these nodes characteristics up to the payoff depth $D$.  This thus corresponds to a network on up to $1+L+L(L-1)+\dots+L(L-1)^{D-1}=1+L \sum_{d=1}^D (L-1)^{d-1}$ nodes and is equal to 2 when $L=1$.  The proportion of individuals of each network type is an equilibrium outcome and one would like to verify which parameter values rationalise the type shares in the data as outcomes of a pairwise stable network.

To do this, the article classifies individuals based on which network types they would not reject.  Depending on the preference shocks, a $B$ individual may be content to have a $W$ connection ($f_{BW}+\epsilon_i(W)>0$), but not a $B$ connection ($f_{BB}+\epsilon_i(B)<0$).  In this case, this individual would not have network type $BB$ in equilibrium as this would contradict pairwise stability, but would be content to be $BW$ or, should there be no $W$ individuals to link to, to remain isolated as $B0$.  The authors thus form \emph{preference classes} collecting all types that would be acceptable to an individual with given realisations of the preference shocks.  In this simple example, the preference class for the individual above would be $\{BB, B0\}$.  (Since there are no connections to be dropped from an isolated network type, the isolated type is an element for every preference class.)

As in the empirical games discussion and the example earlier in this subsection, this corresponds to a partition of the space of unobservable shocks, but only for the individual.  Given a distribution for the preference shocks $\epsilon$, one can compute (either analytically or numerically) the probability of each preference class for a given individual at a particular parameter value.  One can then stipulate how individuals in each preference class are allocated to network types.  In the paper, this is done using \emph{allocation parameters} designating the proportion of individuals in a particular preference class that are allocated to a network type.  In the example above, there are four preference classes for a $B$ individual: $H_1 = \{B0\}, H_2=\{B0,BB\}, H_3 = \{B0,BW\}$ and $H_4=\{B0,BB,BW\}$.  Letting $\alpha_H(t)$ denote the allocation proportion of individuals in preference class $H$ to type $t$, the predicted share of $BW$ individuals is given by $\mathbb{P}(H_1|B)\alpha_{H_1}(BW)+\mathbb{P}(H_2|B)\alpha_{H_2}(BW)+\mathbb{P}(H_3|B)\alpha_{H_3}(BW)+\mathbb{P}(H_4|B)\alpha_{H_4}(BW)$ multiplied by the proportion of $B$ individuals in the group.

The key here is to offer restriction on the allocation parameters $\alpha_H(t)$ to be satisfied for a given profile of network type shares to be consistent with pairwise stability.  These restrictions are necessary conditions for pairwise stability in the article.  For example, nodes can only be allocated to network types that pertain to their preference class: if $t \notin H$, $\alpha_H(t)=0$.  This corresponds to the condition that links have to be beneficial to both individuals and would have $\alpha_{H_1}(BW)=\alpha_{H_2}(BW)=0$.  Second, given any pair of network types that could feasibly add a link to each other (i.e., an isolated individual of either characteristic, $B$ or $W$, in the example), the measure of individuals who would prefer to do so must be zero for at least one of the types.  This corresponds to the condition that non-existing links have to be detrimental to at least one of the individuals.  Another way to express this condition is to require that the product of the measure of individuals of one type that would benefit from adding links to individuals of the other type must be zero.  This translates into a quadratic objective function which in equilibrium has to be zero once allocation parameters are adequately chosen.  Finally, the predicted proportions of network types ought to match the observed proportions of types in the ntwork, which in turn defines a set of linear constraints.

Once these are put together, one can express the task of verifying whether a given parameter vector can rationalise the data (observed network type shares) as that of a pairwise stable network as a quadratic programme on the allocation parameters with constraints requiring those parameters to be positive, add up to one and for predicted shares to be matched in the data.  Should the data be rationalised as a pairwise stable network for a given parameter vector, the optimised objective function is zero.  The cardinality of the problem, while still non-trivial, is related to the cardinality of the preferece classes and network types rather than the potential networks on $|\mathcal{N}_g|$ individuals.  If this is done at each putative parameter, one can then collect those parameters that rationalise the data and form the identified set.  Whereas the quadratic programme above is regrettably not a convex one (since the matrix in the quadratic form is not necessaritly positive definite), the article offers simulation evidence for networks as large as $|\mathcal{N}_g|=500$.  \citeasnoun{Anderson_Richards_Shubik_WP19} applies this framework to the analysis of co-authorships in Economics.\footnote{The article also offers a method to account for sampling uncertainty in the type shares.}

One interesting point to note refers to the source of statistical uncertainty here.  Since there are infinitely many nodes, were the network to be completely observed, network type shares would be perfectly measured.  On the other hand, if there is sampling uncertainty in the measurement of network type shares because only a sample of individuals (and their network types) is collected, sampling uncertainty in type shares would transfer to the estimation of structural parameters.   This relates to what is sometimes referred to as a ``design-based'' paradigm in statistics (see \citeasnoun{kolaczyk2009}). The randomness here obtains from the probability ascribed by the survey scheme to the sampling of the various individuals in the network.  The partial observability of the network is nonetheless not uncommon and rather the norm in many settings.\footnote{Since also focussed on subnetworks, the approach in \citeasnoun{Sheng_ECMA2016} may also be adapted to such contexts.} 

\section{Discussion}

The study of network formation models is an active area of research.  While this survey presents a selective collection of recent developments in the field, several research questions of interest remain and provide a brief discussion for some of those in what follows.  

Since networks mediate and have their formation informed by several outcomes of interest, econometric methods articulating both are potentially important in several areas.  Among the studies modelling the formation of networks as well as outcomes influenced by them, see for example \citeasnoun{gi2013} and \citeasnoun{hsiehlee2016} (using a dyadic network formation model and focussed on educational achievement), \citeasnoun{gilleskiezhang2009} (using Bayes Nash network formation and focussed on smoking initiation),  \citeasnoun{Badev_WP18} and \citeasnoun{hsiehleeboucher2019} (using exponential random graph network formation models along the lines of \citeasnoun{Mele_ECMA2017}, focussing on smoking and, for the latter, also academic achievement).\footnote{For an exposition on econometric methods for outcomes modulated by networks, see \citeasnoun{bdf2020} on this volume.  In those cases, the literature tends to implicitly assume econometric exogeneity between the unobservable variables determining outcomes and the econometric errors determining the networks that mediate those.  This obtains, for instance, when the unobservables determining networks and outcomes are unrelated.  If such a scenario is not empirically adequate, an instrumental variable, control function or a model for network formation may provide potential solutions (see, again, \citeasnoun{bdf2020} for more details).}  Multiplicity is not a salient issue in the econometric network formation protocols in those works (or in the econometric system of equations determining outcomes there).  This nonetheless remains a possibility in other (strategic interaction or conditionally specified) econometric models of network formation.  Multiplicity is also possible in nonlinear systems determining outcomes given networks.  In those cases, partial identification in either stage (i.e., network formation or interactions) may be transmitted to other parameters in the model.  This point is illustrated, for example, in the context of an empirical entry-exit game in industrial organization with potentially multiple equilibria by \citeasnoun{cmt2018}.\footnote{Previous versions of this paper were circulated as ``Inference on Market Power in Markets with Multiple Equilibria.''}  

Relatedly, there is also scope to expand on the catalog of equilibrium notions depending on the protocol for outcome determination and empirical setting of interest. For instance, \citeasnoun{holee2019} (see also \citeasnoun{ghili2018} and \citeasnoun{liebman2018}) articulate network formation and a bargaining framework due to \citeasnoun{hornwolinsky1988} to examine interactions between hospitals and insurers in the United States.  Econometric analysis (and, to a certain extent, theoretical developments) around these and related models remain an area of interest.

Finally, the network itself may be an outcome of interest in a treatment effects context where programmes may lead to changes in the network, which in turn may modulate other outcomes of concern.  \citeasnoun{comolaprina2019}, for example, study the effect of a savings product in Nepal on consumption through a randomised field experiment.\footnote{This article was previously circulated as ``Do Interventions Change the Network? A Dynamic Peer Effect Model Accounting for Network Changes''}  They find that insurance-motivated connections are likely to be rewired after the intervention and taking this into account may improve one's understanding of the programme's effect on consumption.  They offer a treatment response framework on which to analyse this phenomenon and further econometrics work on networks in a potential outcomes setting (possibly also accounting for issues such as multiplicity) would also be a useful area of study.

\newpage

\bibliographystyle{econometrica}
\bibliography{arenetworks_ref}

\end{document}